\documentclass[11pt]{article}
\usepackage{amsmath}
\usepackage{amssymb}
\usepackage{graphicx}

\renewcommand{\baselinestretch}{1.3}
  \renewcommand{\arraystretch}{1.2}
\begin{document}

 \title{Remarks on AKS Primality Testing Algorithm and \\ A Flaw in the Definition of P}

  \author{Zhengjun Cao$^{1,*}$\qquad Lihua Liu$^2$ \\
{\small  $^1$Department of Mathematics, Shanghai University,
  China.   $^*$\,\textsf{caozhj@shu.edu.cn} }\\
{\small $^2$Department of Mathematics, Shanghai Maritime University, Shanghai,
  China.} \\   }

 \date{}\maketitle

\begin{abstract}
We remark that the AKS primality testing algorithm  [Annals of Mathematics 160 (2), 2004]
    needs about 1,000,000,000 G (gigabyte)
   storage space for a number  of 1024 bits. The requirement is very hard to meet.
 The complexity class  P which contains all decision problems
that can be solved by a deterministic Turing machine using a polynomial amount of computation time, is generally believed to be ``easy".
We point out that the time is  estimated only in terms of the amount of arithmetic operations.
It does not comprise \emph{the time for reading and writing data on the tape in a Turing machine}. The flaw
makes some deterministic polynomial time algorithms impractical, and humbles the importance of P=NP question.


 \end{abstract}

\section{Introduction}

The AKS algorithm is a deterministic primality-proving algorithm created by M. Agrawal, N. Kayal, and N. Saxena  \cite{AKS}.  It has attracted much attention in the past decade.
The authors received the 2006 G\"{o}del Prize and the 2006 Fulkerson Prize for this work.
 It is the first primality-proving algorithm to be simultaneously general, polynomial, and deterministic.
  Many fast primality tests  work only for numbers with certain properties. For example, the Lucas-Lehmer test  works only for Mersenne numbers. The  elliptic curve primality test  and Adleman-Pomerance-Rumely primality test \cite{APR}  prove or disprove that a given number is prime,
   but are not known to have polynomial time bounds for all inputs.   Randomized tests, such as Miller-Rabin \cite{Mil76,Rab80}, can test any given number for primality in polynomial time, but are known to produce only a probabilistic result. The correctness of AKS is not conditional on any subsidiary unproven hypothesis. In contrast, the Miller-Rabin test is fully deterministic and runs
 in polynomial time over all inputs, but its correctness depends on the truth of the yet-unproven generalized Riemann hypothesis.

 The AKS algorithm has to find a suitable value $r=O(\mbox{log}^5\,n)$.
 It then checks that $(X+a)^n=X^n+a\, (\mbox{mod}\, X^r-1, n)$ for $a=1$ to $\lfloor\sqrt{\phi(r)} \mbox{log}\, n \rfloor$.  Since $X$ is a free
 variable which is never substituted by a number,
it  has to reduce $(X+a)^n$ in the ring $ \mathbb{Z}_n[X]/(X^r-1)$.
    By the above estimation $r=O(\mbox{log}^5\,n)$, we find that the AKS algorithm  needs about 1,000,000,000 G storage space for a number  of 1024 bits. The requirement is very hard to meet.
   Even worse, it is impossible for current operating systems to write and read data in so huge storage space.  As the authors \cite{AKS} pointed out that $r=O(\mbox{log}^2\,n)$ or $r=O^{\sim}(\mbox{log}^2\,n)$
   if the Artin's conjecture or Sophie-Germain prime density conjecture can be proved.
   In such case, the storage space  required is about 1 G.  If these two conjectures cannot be proved, then
  ``the AKS method is purely academic"  [D. Knuth's comment, in a personal letter].

A Turing machine is a suppositional device that operates symbols on a strip of tape according to a table of rules. It is always assumed that the machine is supplied with \emph{as much tape as it needs}. The time for reading and writing data on the tape in a Turing machine is always
neglected.  We shall remark that the flaw in the definition of P  makes some deterministic polynomial time algorithms impractical.

\section{Description of AKS algorithm}

\begin{center}
\begin{tabular}{|l|}
  \hline
  AKS algorithm\\ \hline
   Input: integer $n>1$.\\
   1. If ($n=a^b$ for $a\in  \mathbb{N} $ and $b>1$), Out COMPOSITE.\\
   2.  Find the smallest integer $r$ such that  $\mbox{ord}_r(n)>\mbox{log}^2\,n$. \\
   3. If $1< \mbox{gcd}\,(a, n) <n$ for some $a\leq r$, output COMPOSITE. \\
   4. If $n\leq r$, output PRIME.\\
   5. For $a=1$ to $\lfloor\sqrt{\phi(r)} \mbox{log}\, n \rfloor$ do\\
   \qquad \quad if ($(X+a)^n\neq X^n+a\, (\mbox{mod}\, X^r-1, n) $), output COMPOSITE.\\
   6. Output PRIME.
    \\   \hline
\end{tabular}\end{center} \vspace*{-1mm}

 The AKS primality test is based upon the following theorem: an integer $n (\geq 2)$ is prime if and only if the polynomial congruence relation
$$(X+a)^n=X^n+a\, (\mbox{mod}\, n)$$
holds for all integers $a$ coprime to $n$. In 2002, the authors \cite{AKS} proved that for appropriately chosen $r$ if the following equation
 $$(X+a)^n=X^n+a\, (\mbox{mod}\, X^r-1, \  n)  \eqno(1)$$
 is satisfied for several $a$'s then $n$ must be a prime power. The number of $a$'s and the appropriate $r$ are both bounded by a polynomial in $\log n$.
   Notice that $r$ is essential to AKS algorithm. It has been proven that $r=O(\mbox{log}^5\,n)$.
We refer to the  Graph-1 for the basic idea behind the proof for the AKS algorithm.

\hspace*{-28mm}\begin{minipage}{\linewidth}
\includegraphics[angle=0,height=20cm,width=19cm]{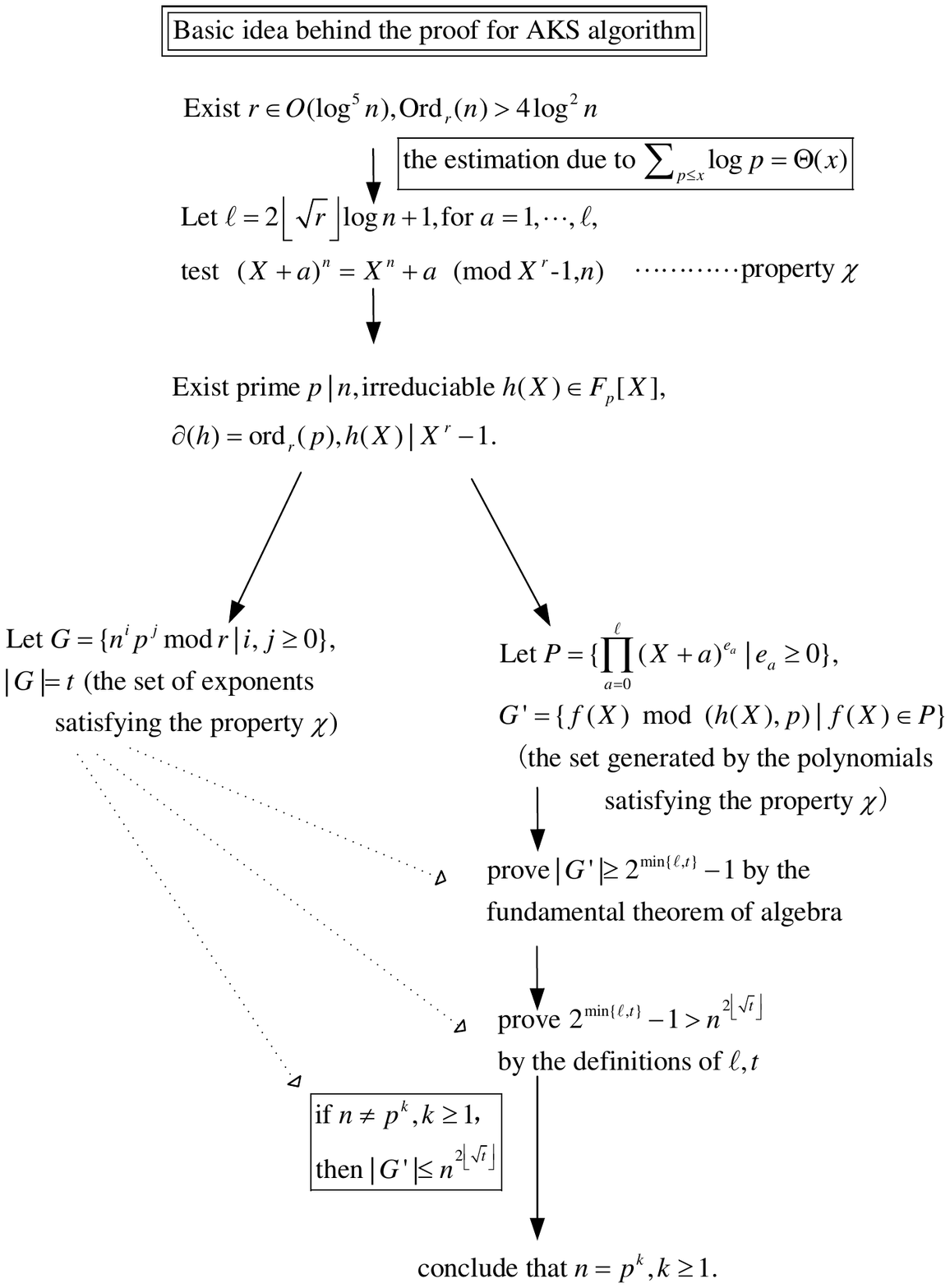}
 \end{minipage}
  \vspace*{-54mm}

  \centerline{Graph-1}

   \section{On the storage requirement for AKS algorithm}

  We note that   Eq.(1) is a polynomial congruence relation where $X$ is a free variable.
   It is never substituted by a number, instead it has to reduce $(X+a)^n$
   in the ring $ \mathbb{Z}_n[X]/(X^r-1)$ and compare the coefficients of the $X$ powers.

   Now suppose that  $n$ is of 1024 bits and $r=  1024^5 k$ where $k$ is a positive number. In the reduction process of $(X+a)^n$ in the ring $ \mathbb{Z}_n[X]/(X^r-1)$,
the AKS algorithm has to compute
$$(c_{r-1}X^{r-1}+\cdots +c_1X +c_0)^2\ (\mbox{mod}\, X^r-1, n), \quad c_i\in \mathbb{Z}_n,\  i=1, \cdots, r-1.  \eqno(2) $$
if it uses  the repeated-squaring method, and store a polynomial of degree  $2r-2$,
$$d_{2r-2}X^{2r-2}+\cdots +d_1X +d_0, \quad d_i\in \mathbb{Z}_n,\  i=1, \cdots, 2r-2.  \eqno(3) $$
Each of these coefficients takes $1024$ bits storage space. Thus the algorithm takes
almost $2\times 1024^6 k$ bits storage space in theory, i.e., about $1024^2$ TG (terabyte).

As of 2013, the storage space for PC is  less than 1 TG. That is to say, the AKS algorithm  needs about 1,000,000 PC's storage space.  To the best of our knowledge, it is impossible for current operating systems to write and read data in so huge storage space.

It has been shown that the best possible estimation for $r$ is $O(\mbox{log}^2\,n)$ if Aritn's Conjecture holds.
In such case, the storage space  required is about 1 GB (gigabyte). In contrast, the Miller-Rabin test only requires several kilobyte (KB) since it is simply computing
$\alpha^{2^s t}\ \mbox{mod}\, n $ for some randomly picked $\alpha\in \mathbb{Z}_n^{+}$, where $2^s t\,|\, n-1, \mbox{gcd}\,(2, t)=1$.

D. Knuth introduced the AKS algorithm in his book (see page 396, The Art of Computer Programming Volume 2: Seminumerical Algorithms, Third Edition).
 We reported to him the flaw in the AKS algorithm. In his reply, he wrote \cite{Knuth13}:
 \begin{quotation}

\emph{Indeed, there are tons of ``polynomial time" algorithms in the
literature that could not possibly be implemented. The
upper bounds that I state on page 396, and the fact that
I don't actually have any exercises or discussion regarding
Theorem A, make it clear that I'm not recommending the
method to programmers.}

\emph{(I have further remarks about the disconnect between
``efficient" and ``polynomial time" on page vii of the preface
to Volume 4A.)}

\emph{You are right that most of the literature about asymptotic
efficiency is off track. I do not have time, however, to
repeatedly say ``don't actually use this". In this particular
case, stating the fourth-power bound should be warning enough,
although I suppose I could have remarked that both
space and time are huge. I kept my discussion as brief
as I could, since I personally find the AKS method
to be interesting but purely academic.  Papers like yours will help to right the balance, I hope.}

 \end{quotation}

\section{On some implementations of the AKS Algorithm}

Some packages for the AKS algorithm can be found at\\
\hspace*{3cm}  \verb|http://fatphil.org/maths/AKS/#Implementations| \\
   Some of them do not check that
$\mbox{ord}_r(n)>\mbox{log}^2\,n$ which is necessary for the correctness of AKS algorithm.

There is a  latest implementation report for AKS algorithm  \cite{LH07}.  For $n=100000000000031$, the chosen number for $r$ is 1024 (see Ref.\cite{LH07}, page 55).
Note that the bit-length of $100000000000031$ is $47$,  $\mbox{ord}_r(n)=32<47^2$.  That is, the chosen number for $r$ is not suitable.  Thus, the estimated running time is not convincing.

\section{Does P mean ``easy"}

   A section in the web page \verb|http://en.wikipedia.org/wiki/P_versus_NP_problem|, is devoted to
 the problem wether P means ``easy".
 It emphasizes that: 1) a theoretical polynomial algorithm may have extremely large constant factors or exponents thus rendering it impractical, 2)
 there are types of computations which do not conform to the Turing machine model on which P and NP are defined.

Recall that a Turing machine is a suppositional device that operates symbols on a strip of tape according to a table of rules.
It is always assumed that the machine is supplied with as much tape as it needs.  In complexity theory, P is the class which contains all decision problems
that can be solved by a deterministic Turing machine using a polynomial amount of computation time.
The time is estimated only in terms of the amount of arithmetic operations.
It does not comprise the time for reading and writing data on the tape in a Turing machine. The flaw in the definition of P is not generally known to the public
(refer to the above web page).
From the practical point of view, it is definite that some
deterministic polynomial time algorithms with the requirement of huge storage space can not be successfully implemented.
In some senses, the flaw greatly humbles the importance of P=NP question.

Here is a good collection of papers or works that try to settle the ``P versus NP" question \\
 \hspace*{30mm}\verb|http://www.win.tue.nl/~gwoegi/P-versus-NP.htm|\\
 Amusingly, the gap between approval and disapproval is negligible.

\section{Conclusion}
In this paper,  we remark that the AKS algorithm is flawed if the related conjectures can not be proved. We also remark that the running time of an algorithm should not be estimated simply in terms of
the amount of arithmetic operations.

\end{document}